\documentclass[useAMS,usenatbib]{mn2e}

\usepackage{times,graphicx,amssymb,rotating}

\title[Photoionisation rates around T Tauri Stars]{On the origin of ionising photons emitted by T Tauri stars}

\author[R.D.~Alexander, C.J.~Clarke \& J.E.~Pringle]
  {R.D.~Alexander\thanks{email: rda@ast.cam.ac.uk},
  C.J.~Clarke
  and J.E.~Pringle \\
  Institute of Astronomy, Madingley Road, Cambridge, CB3 0HA}

\begin{document}

\pagerange{\pageref{firstpage}--\pageref{lastpage}} \pubyear{2003}

\maketitle

\label{firstpage}

\begin{abstract}
We address the issue of the production of Lyman continuum photons by T Tauri stars, in an attempt to provide constraints on theoretical models of disc photoionisation.  By treating the accretion shock as a hotspot on the stellar surface we show that Lyman continuum photons are produced at a rate approximately three orders of magnitude lower than that produced by a corresponding black body, and that a strong Lyman continuum is only emitted for high mass accretion rates.  When our models are extended to include a column of material accreting on to the hotspot we find that the accretion column is extremely optically thick to Lyman continuum photons.  Further, we find that radiative recombination of hydrogen atoms within the column is not an efficient means of producing photons with energies greater than 13.6eV, and find that an accretion column of any conceivable height suppresses the emission of Lyman continuum photons to a level below or comparable to that expected from the stellar photosphere.  The photospheric Lyman continuum is itself much too weak to affect disc evolution significantly, and we find that the Lyman continuum emitted by an accretion shock is similarly unable to influence disc evolution significantly.  This result has important consequences for models which use photoionisation as a mechanism to drive the dispersal of circumstellar discs, essentially proving that an additional source of Lyman continuum photons must exist if disc photoionisation is to be significant.
\end{abstract}



\section{Introduction}
The evolution and eventual dispersal of discs around young stars is an important area of study, as it provides constraints on theories of both star and planet formation.  It is now well established that the majority of young stars at ages of $\sim 10^6$yr have circumstellar discs which are optically thick at optical and infrared wavelengths \citep*{strom89,kh95,haisch01}.  These discs are relatively massive, with masses of the order of a few percent of a solar mass \citep*{beckwith90,ec03}.  However by an age of $\sim 10^7$yr most stars are no longer seen to have such massive discs, although low-mass ``debris discs'' may remain (eg.~\citealt*{ms97,wyatt03}).  The mechanism by which these discs are dispersed remains an important unsolved question.

One fact which is clear, however, is that a large fraction of the mass from the disc will eventually be accreted on to the central (proto)star.
The inner edge of the accretion disc is typically truncated by the magnetosphere at a radius of around 5$R_*$ \citep*{mch97,cg98} and so material falling from the disc on to the central star can attain an extremely high velocity, resulting in a so-called ``accretion shock'' when this material impacts upon the stellar surface.  Existing models of the accretion shock \citep*{cg98,gul98,lamzin98,gul00} have paid a great deal of attention to the emission in the ultraviolet (1000-3000\AA) and visible (3500-7000\AA) wavebands, comparing theoretical predictions to observed spectra.  Extremely good models have been constructed, and these emission spectra are now well understood.  However, very little attention has been paid to the emission shortward of the Lyman break ($<$ 912\AA), primarily because absorption by interstellar H{\sc i} makes it impossible to observe young stars in this wavelength regime.  Recent theoretical studies \citep*{cc01,mjh03,acp03} have suggested that photoionisation by the central object may play an important role in disc dispersal, and so the emission shortward of the Lyman break has become important.  

Currently the origin of photoionising radiation from young stars such as T Tauri stars is unclear, and even the magnitude of such emission is poorly constrained \citep{gahm79,ia87}.  To date, models of disc photoionisation have used either a constant ionising flux (eg.~\citealt{holl94,cc01}), assumed to be chromospheric in origin, or modelled the accretion-driven flux simply as a constant temperature hotspot on the stellar surface, emitting as a blackbody \citep{mjh03}.  The latter produces an ionising flux which is proportional to the mass accretion rate, thus decreasing dramatically with time as the accretion rate falls, and so naturally the models of \citet{cc01} and \citet{mjh03} have produced markedly different results.  These models \citep*{holl94,ry97,hollppiv} also find that the mass-loss rate scales approximately with the square root of the ionising flux, and so very high ionising fluxes are required to influence disc evolution significantly: an ionising flux of $10^{41}$photons s$^{-1}$ will drive mass-loss from the disc at around the $10^{-10}$$M_{\odot}\mathrm{yr}^{-1}$ level.

In this paper we study the issue of the ionising flux generated by an accretion shock.  The simplified model of the accretion shock adopted by \citet{mjh03} neglects two key points, which we address in turn.  Firstly, it seems likely that the Lyman continuum emission from such a hotspot would resemble a stellar atmosphere rather than a blackbody; whilst the two are almost identical at longer wavelengths, photo-absorption by H{\sc i} provides a strong suppression of the flux shortward of 912\AA.  Secondly, the photons emitted by the accretion shock must pass {\it through} the column of material accreting on to it in order to interact with material in the disc, and again we would expect photo-absorption by H{\sc i} in the column to supress the Lyman continuum significantly.  In order to address these issues we have modelled this process in some detail.  In section \ref{sec:atmos} we investigate the effect of replacing the blackbody hotspot with a more realistic stellar atmosphere.  In section \ref{sec:columns} we investigate the effect of passing these photons, from both the blackbody and the stellar atmosphere, through a column of accreting material.  In section \ref{sec:dis} we discuss our results and the limitations of our analysis, and in section \ref{sec:summary} we summarise our conclusions.


\section{Stellar atmospheres}\label{sec:atmos}
The photoionisation models of \citet{mjh03} model the ionising photons as follows.  They assume that the flux from the accretion shock can be modelled as a constant temperature hotspot, and adopt a blackbody spectral energy distribution at a temperature of $T=15,000$K.  They assume that half the accretion luminosity is radiated by this hotspot, and so for a star of mass $M_*$ and radius $R_*$ the accretion shock luminosity $L$ is given by:
\begin{equation}\label{eq:mjh}
L = \frac{G M_* \dot{M}_{\mathrm {d}}}{2R_*} = A\sigma_{\mathrm {SB}}T^{4}
\end{equation}
where $\dot{M}_{\mathrm {d}}$ is the rate of mass accretion from the disc, $A$ is the area of the hotspot, and $\sigma_{\mathrm {SB}}$ is the Stefan-Boltzmann constant.  As the temperature is constant the rate of ionising photons $\Phi_{\mathrm {a}}$ is simply proportional to the accretion rate $\dot{M}_{\mathrm {d}}$.  In addition, \citet{mjh03} add a further contribution to the ionising flux from the stellar photosphere, neglecting the poorly-constrained chromospheric contribution, at a constant rate of $\Phi_{\mathrm {p}}=1.29\times10^{31}$photons s$^{-1}$, with the total ionising photon rate given by $\Phi_{\mathrm {a}}+\Phi_{\mathrm {p}}$.  Our assertion is that, given the high density of atomic hydrogen, it is extremely unlikely that such a hotspot would radiate as a blackbody.  A spectrum akin to a stellar atmosphere, showing a significant ``Lyman edge'', seems much more likely.  As a result, we first consider the strength of this effect.

\subsection{Constant temperature}\label{sec:const_t}
\begin{figure}
        \resizebox{\hsize}{!}{
        \begin{turn}{270}
        \includegraphics{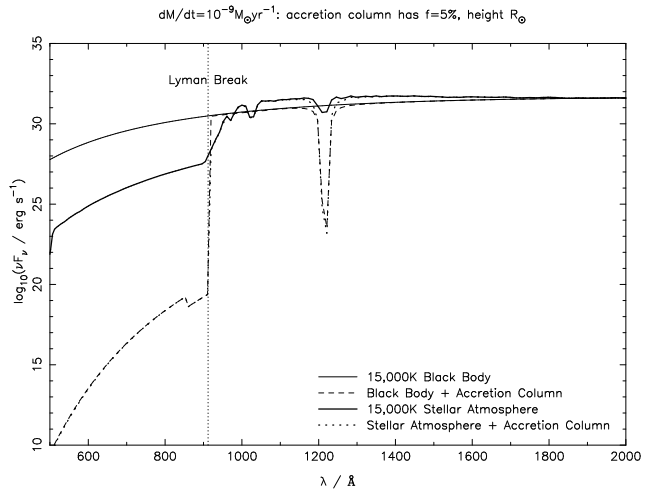}
        \end{turn}
        }
        \caption{Spectra of light incident on and emitted by a constant density accretion column with a covering factor of 5\% and a height of 1R$_{\odot}$.  Note the precipitous drop in the emitted spectra at the Lyman break, due to photoabsorption by H{\sc i}.  The deep absorption feature at 1215\AA~is Ly$\alpha$.}
        \label{fig:spectra}
\end{figure}

\begin{figure}
        \resizebox{\hsize}{!}{
        \begin{turn}{270}
        \includegraphics{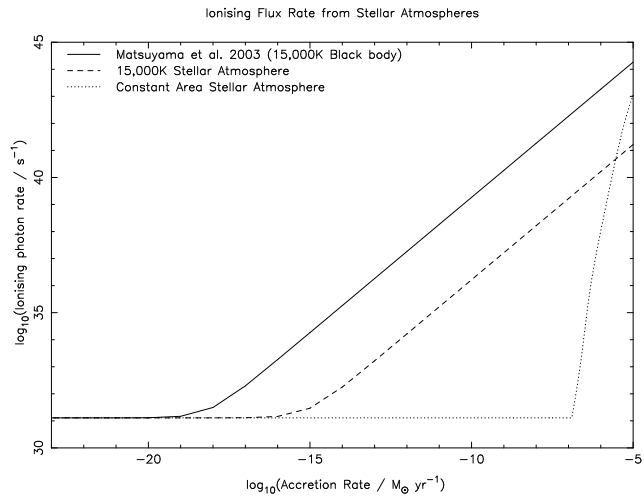}
        \end{turn}
        }
        \caption{Ionising flux rates from different accretion shock models.  The rate from a simple model atmosphere is some 3 orders of magnitude less than that from a blackbody, and the constant area case, with $T \propto \dot{M}_{\mathrm {d}}^{1/4}$ decays more precipitously still.}
        \label{fig:atmos}
\end{figure}

Our first models simply involve substituting model stellar atmospheres in place of the blackbody emission in equation \ref{eq:mjh}.  We have adopted the same stellar parameters as \cite{mjh03} ($R=1R_{\odot}$, $M=1M_{\odot}$), and similarly adopted a constant hotspot temperature of 15,000K.  The luminosity $L$ scales with $\dot{M}_{\mathrm {d}}$ in the same manner as in equation \ref{eq:mjh}.  We have utilised the Kurucz model atmospheres \citep{kurucz92} (which have been incorporated into the {\sc cloudy} code) for this temperature and surface gravity.  The model atmospheres do not deviate significantly from the blackbody at longer wavelengths, but are some 3 orders of magnitude less luminous than the corresponding blackbody at wavelengths shortward of the Lyman limit at 912\AA, due to absorption by H{\sc i}.  Fig.~\ref{fig:spectra} compares the blackbody and Kurucz spectra.  (The apparent lack of emission lines in Fig \ref{fig:spectra} is an artefact of the relatively large bin-width used within the code.  The use of such large wavelength bins results in line fluxes which are negligible in comparison to the continuum.)  Fig.\ref{fig:atmos} plots the ionising fluxes as a function of accretion rate, and we see that the stellar atmosphere hotspot produces ionising photons at a rate that is a factor of 1100 less than that obtained from the blackbody model.

\subsection{Constant area}
Another consideration is that of the hotspot area.  The \citet{mjh03} blackbody formulation described in equation \ref{eq:mjh} uses a hotspot temperature which remains constant for different mass accretion rates, implying a hotspot size which decreases as the accretion rate drops.  The accreting material is thought to be channelled on to the magnetic poles as it falls on to the stellar surface, and unless the topology of the magnetic field varies systematically wth the accretion rate, the hotspot area should remain approximately constant with time.  If the hotspot area $A$ is constant then we would expect, from equation \ref{eq:mjh}, the hotspot temperature to vary as $T \propto \dot{M}_{\mathrm {d}}^{1/4}$ The {\it total} luminosity of a model atmosphere is very similar to that of a blackbody, and so we adopt this relationship for the stellar atmospheres also.  We re-evaluated the model atmospheres described in section \ref{sec:const_t}, keeping the scaling luminosity proportional to $\dot{M}_{\mathrm {d}}$, but now with the temperature given by:
\begin{equation}\label{eq:const_a}
T = \left(\frac{GM_*}{2R_* A \sigma_{\mathrm {SB}}} \dot{M}_{\mathrm {d}} \right)^{1/4}
\end{equation}
The result of this is shown in Fig.\ref{fig:atmos}; the drop-off in the ionising photon rate is much more precipitous than in the constant temperature case, with only very high mass accretion rates, greater than $10^{-7}$$M_{\odot}$yr$^{-1}$, producing ionising photons at greater than the photospheric rate.  
In fact, the relationship between the hotspot area and the mass accretion rate is not well understood, and \citet{cg98} even found observational evidence for a hotspot area which increases with $\dot{M}_{\mathrm {d}}$ (which would imply an even steeper decline in the ionising flux).  However, we go on to show that the presence of an accretion column above the hotspot is by far the dominant factor in controlling the ionising photon rate, and so the exact details of the hotspot area are not of great significance.


\section{Accretion Columns}\label{sec:columns}
The other issue which affects the emitted ionising flux is the assumed presence of a column of accreting material directly above the hotspot.  This material will absorb Lyman continuum photons through photoionisation of H{\sc i}, and so a large attenuation of the ionising flux is expected.  Adopting the photoionisation cross-section from \citet{allen} of $\sigma_{\mathrm {13.6eV}} = 6.3$$\times$$10^{-18}$cm$^2$, indicates that any column density greater than 5$\times$$10^{18}$cm$^{-2}$ will result in an attenuation of the incident flux by a factor of $>10^{13}$, enough to reduce {\it any} incident ionising photon rate to below photospheric levels.  The density of the infalling material is of order 5$\times$$10^{12}$cm$^{-3}$ \citep{cg98} and so this results in an attenuation length of order 10$^{6}$cm (10$^{-5}$$R_{\odot}$).  As a result, the only Lyman continuum photons which can be emitted, at any significant rate, by an accretion column must be due to radiative recombination of hydrogen in the column, the so-called diffuse continuum.

In order to investigate this effect further we have constructed models of the accretion column using the {\sc cloudy} photoionisation code \citep{cloudy}.  These models consist of a uniform central source, radiating either as a blackbody or a stellar atmosphere, and an accretion column.  The central source emits in the radial direction only, and the accretion column covers a constant solid angle.  Thus, by a simple linear subtraction of the emission not incident on the column we can treat the column as if it were illuminated solely by a hotspot at its base with a flux in the radial direction.  In reality such a hotspot would produce some lateral component of flux near to its edges: this is discussed in section \ref{sec:dis}.  The accretion column has a solar chemical composition (although as the dominant effect is photoionisation of H{\sc i} there is almost no dependence on chemical composition) and covers a small fraction $f$ of the stellar surface.  Following \citet{cg98} we adopt a ``free-fall'' scaling density of the form:
\begin{eqnarray}\label{eq:hden}
n_{\mathrm H}(0) = 5.2\times 10^{12}\mathrm {cm}^{-3} \left(\frac{\dot{M}_{\mathrm {d}}}{10^{-8}M_{\odot}\mathrm{yr}^{-1}}\right) \left(\frac{M_*}{0.5 M_{\odot}}\right)^{-1/2}
\nonumber \\
\times \left(\frac{R_*}{2 R_{\odot}}\right)^{-3/2} \left(\frac{f}{0.01}\right)^{-1}
\end{eqnarray}
The radial behaviour of the density is obtained by assuming that the material falls along magnetic field lines, in a manner consistent with standard magnetospheric accretion models (eg.~\citealt{gl78}).  Therefore at a given radius the product of the field strength and the column cross-sectional area is a constant.  Assuming a dipole magnetic field we have $B \propto R^{-3}$, and therefore the cross-sectional area is proportional to $R^3$.  For a column with cross-sectional area $A'$ and mass density $\rho$, mass conservation requires that:
\begin{equation}
\rho A' v_{\mathrm {ff}} = \mathrm {constant}
\end{equation}
The number density $n_{\mathrm H} \propto \rho$, and the free-fall velocity $v_{\mathrm {ff}} \propto R^{-1/2}$, and so we have a density scaling law of:
\begin{equation}
n_{\mathrm H} \propto R^{-5/2}
\end{equation}
with the condition in equation \ref{eq:hden} used to fix the scaling constant.

{\sc cloudy} allows us to evaluate the continuum and line emission from the top of the accretion column, which is a combination of the continuum incident on the bottom of the column, attenuated by the column, and the diffuse emission from the heated column.  It does not allow direct evaluation of the emission from the ``sides'' of the accretion column, which may be significant and is discussed in section \ref{sec:dis}. 

Again, we have adopted $R=1R_{\odot}$ and $M=1M_{\odot}$, and have constructed models of these accretion columns for a broad range of accretion rates, hotspot areas and column heights.  Initially both the blackbody and stellar atmosphere hotspot formulations were used, with the ``constant temperature'' formalism used as it provides the greatest ionising flux to the column and can be treated as a limiting case.  However, as discussed above, the incident Lyman continua from both are extinguished over a very short length scale, and so the only Lyman continuum emission which emerges from the columns is the so-called ``diffuse'' emission due to the radiative recombination of atomic hydrogen.  As seen in Fig.\ref{fig:spectra}, the Lyman continua emitted by the columns are identical in both cases, and depend only the nature of the column rather than the spectrum of the illuminating hotspot, as both hotspot spectra have the same bolometric luminosity.  As a result, only the more realistic stellar atmosphere models were used for the remainder of the cases.

\subsection{Results}\label{results}
\begin{figure}
        \resizebox{\hsize}{!}{
        \begin{turn}{270}
        \includegraphics{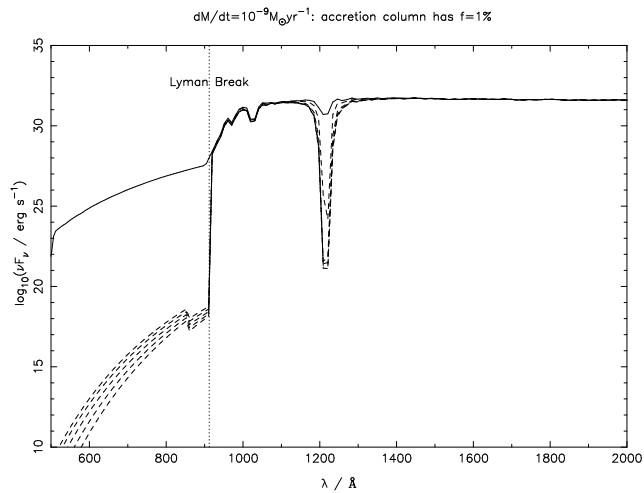}
        \end{turn}
        }
        \caption{Incident (solid line) and transmitted (dashed lines) spectra for model accretion column with $\dot{M}_{\mathrm {d}}=10^{-9}M_{\odot}\mathrm{yr}^{-1}$ and a covering fraction of 1\%.  The spectra shown are for column heights of 0.2, 1.0, 2.0, 3.0 and 4.0$R_{\odot}$  (brightest to faintest respectively).}
        \label{fig:colspec}
\end{figure}

\begin{figure}
        \resizebox{\hsize}{!}{
        \begin{turn}{270}
        \includegraphics{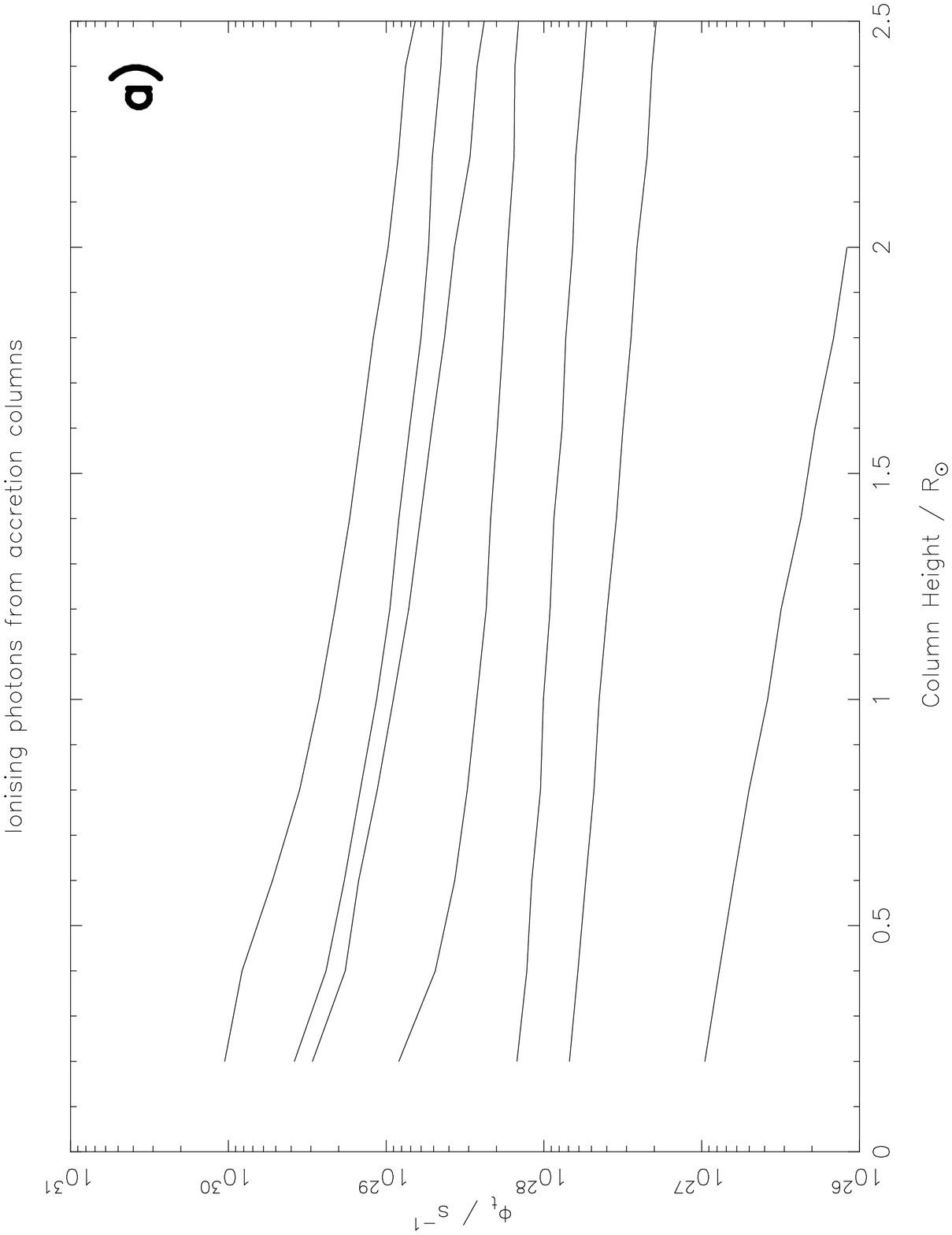}
        \end{turn}
        }

	\vspace*{4mm}

        \resizebox{\hsize}{!}{
        \begin{turn}{270}
        \includegraphics{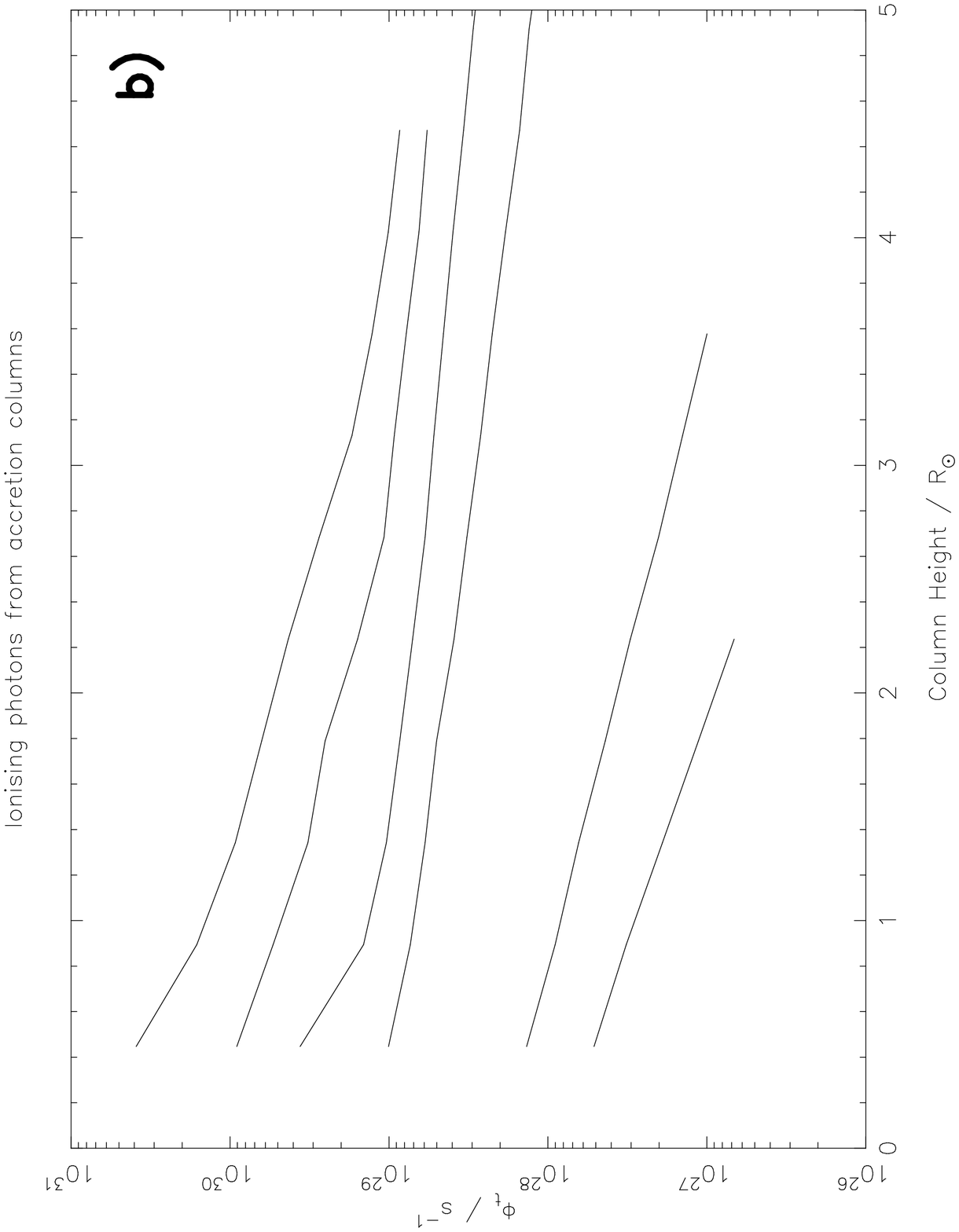}
        \end{turn}
        }

	\vspace*{4mm}

        \resizebox{\hsize}{!}{
        \begin{turn}{270}
        \includegraphics{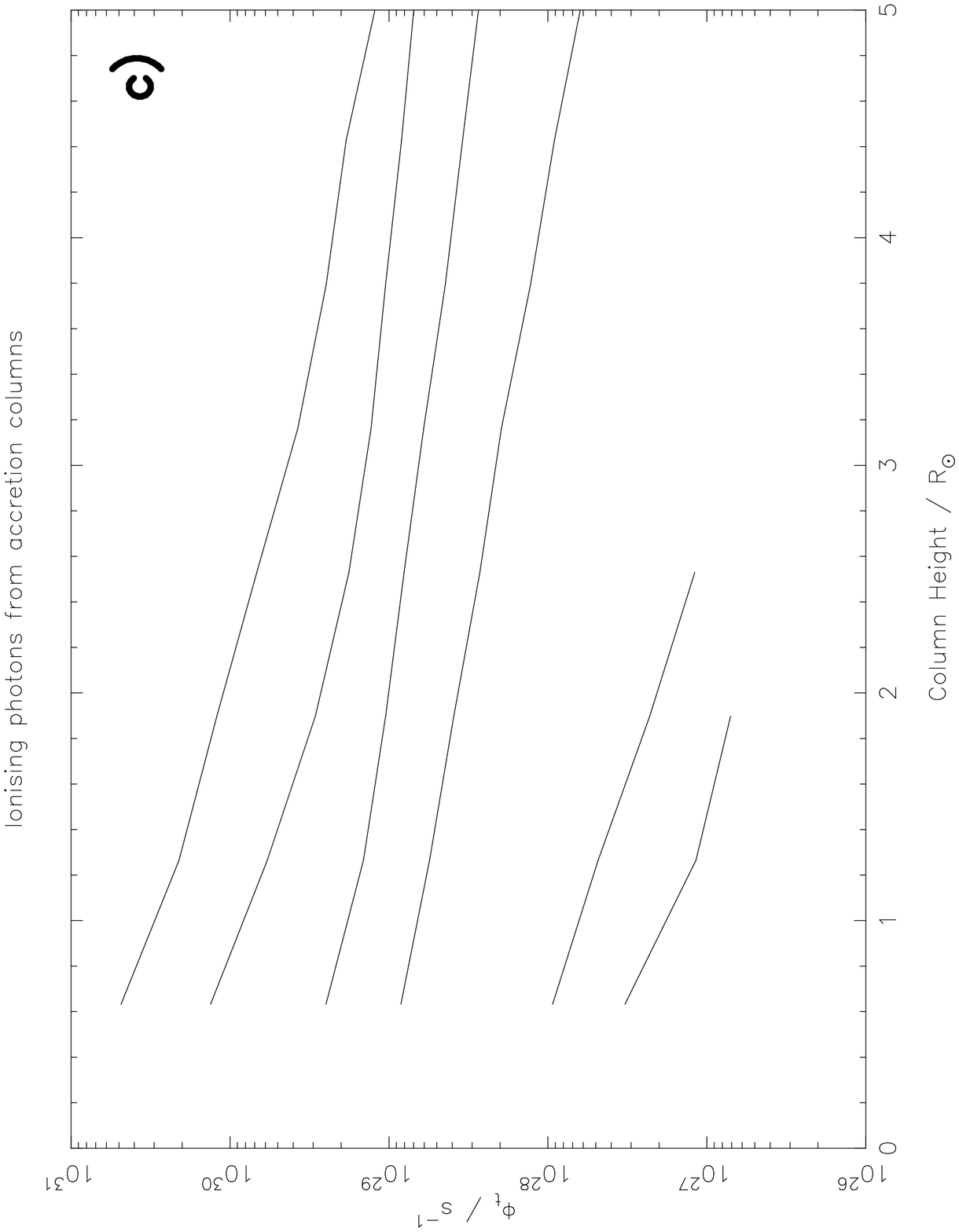}
        \end{turn}
        }
        \caption{Ionising photon rates emitted by accretion columns of different heights, for columns with covering factors (and hotspot areas) of: {\bf a)} 1\%; {\bf b)} 5\%; {\bf c)} 10\% .  From top to bottom in each plot the curves are for accretion rates of $1\times10^{-7}$, $5\times10^{-8}$, $1\times10^{-8}$, $5\times10^{-9}$, $1\times10^{-9}$, $5\times10^{-10}$ \& $1\times10^{-10}$$M_{\odot}\mathrm{yr}^{-1}$ respectively.  The smallest column height in each case was taken to be half the hotspot radius.  Where the curves stop abruptly it is because the temperature in the column fell below the limiting temperature of 3000K, and for the same reason the last curves ($\dot{M}_{\mathrm {d}} = 1\times10^{-10}$$M_{\odot}\mathrm{yr}^{-1}$) in plots {\bf b)} and {\bf c)} have been omitted completely.}
        \label{fig:phi}
\end{figure}

The results of the simulations described above are presented in Figs.\ref{fig:colspec} \& \ref{fig:phi}.  Fig.\ref{fig:colspec} shows the incident and transmitted spectra for a typical case at a variety of different column heights, and Fig.\ref{fig:phi} shows the dependence of the emergent ionising flux on column height.  {\sc cloudy} is only valid for temperatures greater than $\simeq$3000K - for temperatures lower than this the thermal solutions are no longer unique - and for large column heights or cases with little heating (ie.~low accretion rates) the temperature dropped below this critical value.   The models were not pursued beyond this point, with 3000K used as the temperature limit of the calculations.  As seen in Fig.\ref{fig:phi}, the emergent ionising flux, as expected, decreases with both decreasing accretion rate and with increasing column height.  Smaller hotspots result in higher density of material in the accretion column for a given accretion rate, and so tend to produce slightly larger ionising fluxes.  However these variations are small relative to those due to variations in column height or accretion rate.

The most important point, however, is that the emergent ionising photon rates for {\it all} of the columns are less than the photospheric value of 10$^{31}$photons s$^{-1}$.  Further, the photospheric value is itself some 10 orders of magnitude lower than the rate required to influence disc evolution significantly.  This means that the accretion columns we have modelled cannot emit ionising photons at a rate that will be significant in disc photoionisation models, for any choice of parameters.  


\section{Discussion}\label{sec:dis}
There are obvious caveats to these models.  The first, and most significant, is that the {\sc cloudy} code is only completely reliable at densities less than $10^{13}$cm$^{-3}$; it is prone to numerical problems at higher densities, and for cases of high $\dot{M}_{\mathrm {d}}$ and small $f$ the density in our models can exceed this value.  However, the main uncertainties at these densities are regarding the treatment of heavy elements.  The dominant effects in the regime in which we are interested are photoionisation and recombination of atomic hydrogen, and these processes {\it are} treated reliably by the code.  However in order to compensate for this numerical problem we were forced to limit the density artificially to have a maximum value of $5\times10^{13}$cm$^{-3}$.  This affected the 3 models with the highest densities ($\dot{M}_{\mathrm {d}} = 1\times10^{-7}$$M_{\odot}\mathrm{yr}^{-1}$, $f=0.01,0.05$ and $\dot{M}_{\mathrm {d}} = 5\times10^{-8}$$M_{\odot}\mathrm{yr}^{-1}$, $f=0.01$), and the reduction in density results in these models under-estimating the ionising flux somewhat.  This does introduce some uncertainty into our models, but we consider the effect to be small relative to the gross effects which dominate the calculations.

A further caveat regards the issue of non-radial emission, both from the hotspot and from the ``sides'' of the column, as mentioned in section \ref{sec:columns}.  The ionising photon rates from our models are those emitted from the top of the accretion columns only, and neglect emission from the sides of the columns.  Further, the incident flux from the hotspot is assumed to be purely radial.  The Lyman continuum emission in which we are interested arises from the radiative recombination of hydrogen atoms, an intrinsically isotropic emission process, and so emission from the sides of the column could be significant.  However, as the recombination process is isotropic it is reasonable to assume that the Lyman continuum emitted from the sides of a column will be comparable to that emitted from the top of a column that is truncated at a height equal to its diameter.  Fig.\ref{fig:phi} shows that the diffuse Lyman continuum from the top of the column decreases dramatically as the column height increases (and the density decreases).  Consequently the Lyman continuum emission from the sides of the column will only be significant over a distance comparable to the hotspot diameter: the sides of the column only produce a significant Lyman continuum near to the stellar surface.  

As noted in section \ref{sec:columns}, Lyman continuum photons incident on the columns will be attenuated by a factor of $10^{13}$ over a distance of approximately 10$^{-5}$$R_{\odot}$.  As a result of this, the majority of the non-radial Lyman continuum photons emitted by an isotropically emitting hotspot will be absorbed by the column.  The only such photons not absorbed will be those emitted within a fraction of an attenuation length of the edge of the hotspot, in a direction away from the centre of the hotspot.  A hotspot covering 1\% of the stellar surface has a radius of 0.2$R_{\odot}$, so less than $10^{-5}$ of the hotspot photons are emitted within 10$^{-5}$$R_{\odot}$ of the hotspot edge.  Given this fact, and also the behaviour of the ``raw'' hotspot emission described in Fig.\ref{fig:atmos}, we neglect this effect.

The net result of these two simplifications is that in reality the emission from the bottom of any accretion column will dominate the Lyman continuum emitted by the column, and accretion columns of any height will emit ionising photons at a rate comparable to that provided by the lower part of the column (the left-hand end of the curves in Fig.\ref{fig:phi}).  This will increase the largest calculated ionising photon rates by a small geometric factor but still cannot increase their flux to significantly greater than 10$^{31}$photons s$^{-1}$; the photospheric emission will still dominate the overall Lyman continuum.  More importantly, much higher ionising fluxes (of the order of 10$^{41}$photons s$^{-1}$) are required to have a significant effect on disc evolution \citep{cc01,mjh03}, and the uncertainties caused by the approximations we have made are negligible compared to this 10 orders-of-magnitude difference.

Further simplifications used in our models regard the geometry of the accretion column.  Our models use a column with a constant covering factor - essentially a truncated radial cone - and so the area of the outer surface at a given radius $R$ is proportional to $R^2$.  However in reality the column is channelled by the magnetosphere and, as explained in section \ref{sec:columns}, has an area proportional to $R^3$; our model accretion columns are somewhat less flared than we would expect to see in reality.  However the difference between the two is only significant at large radii; small column heights, which provide the highest ionising fluxes, will not show significant deviation between the two cases.  Again, the net result of this is that we probably under-estimate the ionising flux by a small factor, but not by enough to alter the results significantly.

As our accretion columns are radial cones they do not bend to follow the magnetosphere, as expected in more realistic magnetospheric accretion models (eg.~\citealt{gl78}).  In such a model the column would bend over to meet the accretion disc, with a curvature dependent on both the latitude of the hotspot and the strength of the magnetosphere.  However, as discussed above, the emission from the bottom part of the accretion column dominates over that from the upper parts (those affected by this curvature), so this simplification will not affect our results significantly.

There is also the issue of an infall velocity, which our models do not address.  In reality the accretion column will be falling towards the stellar surface at close to the free-fall velocity, which can be several hundred kms$^{-1}$, and so this could modify the absorbing effect of the cloud.  However the infall velocity is much less than the speed of light, and there are no strong emission lines near to the Lyman break, and so we consider the impact of this effect on the emitted Lyman continuum to be negligible.

It should be noted that our model makes no predictions as to the behaviour of the photons emitted at wavelengths longward of the Lyman break.  As shown in Fig.\ref{fig:spectra}, the emission from the top of the column longward of the Lyman break is essentially identical to that from the hotspot at the base of the column.  We approximate the accretion shock crudely, and so are not able to fit our models to observed spectra in a manner similar to \citet{cg98} or \citet*{jkvl00}.  However the Lyman continuum emitted by our columns is insensitive to variations in the hotspot spectrum, due to the high optical depth of the columns to Lyman continuum photons.  Consequently we find that any reasonable accretion-shock model will produce a similar Lyman continuum, and that this Lyman continuum is independent of the emission at longer wavelengths.

We have adopted stellar parameters of $R=1R_{\odot}$ and $M=1M_{\odot}$ to provide direct comparisons with the models of \citet{cc01} and \citet{mjh03}.  However in the case of T Tauri stars a radius of $2R_{\odot}$ and a mass of $0.5M_{\odot}$ would be more realistic \citep{gul98}.  The result of this will be that our models over-estimate the ionising flux somewhat, due to both reduction in the energy released by accretion on to the stellar surface and also due to a reduction in the star's surface gravity.  Once again, however, it is unlikely that these factors are significant in comparison to the gross effects we have already considered.  
Similarly, the use of the ``constant temperature'' hotspot to heat the column probably over-estimates both the ionising flux and heating provided by the hotspot, and thus over-estimates the diffuse Lyman continuum.  In effect we have constructed a ``best-case'' model, designed to produce the maximum ionising flux, and still found the ionising flux to be less than that emitted by the stellar photosphere.  It seems extremely unlikely that any conceivable accretion column could produce ionising photons at a rate significantly greater than this.


\section{Summary}\label{sec:summary}
In an attempt to provide some constraints on the nature and magnitude of the ionising continuum emitted by T Tauri stars, we have constructed models which treat the accretion shock as a hotspot on the stellar surface beneath a column of accreting material.  We have modelled these columns for a variety of different accretion rates, hotspot sizes and column heights, and have found that:
\begin{itemize}
\item A hotspot radiating like a stellar atmosphere radiates ionising photons at a rate some 3 orders of magnitude less than the corresponding blackbody.
\item A constant area hotspot radiating like a stellar atmosphere can only emit ionising photons at greater than photospheric rates for mass accretion rates greater than $10^{-7}$$M_{\odot}$yr$^{-1}$.  Such accretion rates are near the upper limit of the rates derived from observations \citep{hcga98,jkvl00}.
\item Photoionisation of neutral hydrogen in the accretion column attenuates the Lyman continuum from any hotspot to zero over a very short length scale.  The ionising photons which do emerge are due to radiative recombination of hydrogen atoms in the column, and the rate of ionising photon emission is less than the photospheric level for all of the accretion columns we have modelled.
\end{itemize}
In short, we find that accretion shocks and columns are extremely unlikely to produce Lyman continuum photons at a rate significantly greater than that expected from the stellar photosphere.  The photospheric level itself is some 10 orders of magnitude below the rates required for photoionisation to affect disc evolution significantly, and so it seems that the Lyman continuum emitted by an accretion shock will not be large enough to be significant in disc photoionisation models.  These models have provided an attractive explanation of some observed disc properties (eg.~\citealt{cc01,acp03}) but we have shown, as suggested by \citet{cc01}, that they must be powered by something other than the accretion-shock emission.  In order to produce ionising photons at a rate large enough to enable photoionisation models to provide a realistic means of disc dispersal, the central objects must sustain a source of ionising photons that is not driven by accretion from the disc.  


\section*{Acknowledgments}
We thank Bob Carswell for useful discussions and advice on the use of the {\sc cloudy} code.  RDA acknowledges the support of a PPARC PhD studentship.  CJC gratefully acknowledges support from the Leverhulme Trust in the form of a Philip Leverhulme Prize.
We thank an anonymous referee for helpful comments which improved the clarity of the paper.


\label{lastpage}

\end{document}